\documentclass[twocolumn,superscriptaddress,amsmath,amssymb,aps,pra]{revtex4}
\usepackage{amsmath,amsthm,amssymb}
\usepackage{latexsym}
\usepackage{amscd,graphicx}
\usepackage[titletoc]{appendix}
\usepackage{epsfig}
\usepackage{epstopdf}
\usepackage[normalem]{ulem}
\usepackage[dvipsnames]{xcolor}
\usepackage[colorlinks=true,linkcolor=red,citecolor=blue]{hyperref}

\newcommand{\abs}[1]{| #1 |}

\begin{document}
	\title{Nonreciprocal Photon-Phonon Entanglement in Kerr-Modified Spinning Cavity Magnomechanics }
	
	\author{Jiaojiao Chen}
	\affiliation{Department of Physics and Optoelectronic Engineering,Anhui University, Anhui 230000, China}
	\affiliation{Department of Physics, Wenzhou University, Zhejiang 325035, China}
	
	\author{Xiao-Gang Fan}
	\affiliation{Department of Physics and Optoelectronic Engineering,Anhui University, Anhui 230000, China}
	
	\author{Wei Xiong}
	\altaffiliation{xiongweiphys@wzu.edu.cn}
	\affiliation{Department of Physics, Wenzhou University, Zhejiang 325035, China}
	\affiliation{Department of Physics and Optoelectronic Engineering,Anhui University, Anhui 230000, China}
	
	\author{Dong Wang}
	\affiliation{Department of Physics and Optoelectronic Engineering,Anhui University, Anhui 230000, China}
	
	\author{Liu Ye}
	\altaffiliation{yeliu@ahu.edu.cn}
	\affiliation{Department of Physics and Optoelectronic Engineering,Anhui University, Anhui 230000, China}

	\date{\today }
	
	\begin{abstract}
Cavity magnomechanics has shown great potential in studying macroscopic quantum effects, especially for quantum entanglement, which is a key resource for quantum information science. Here we propose to realize magnon mediated nonreciprocal photon-phonon entanglement, {which exhibits asymmetry when opposite magnetic or driving fields are respectively applied to the magnons with the Kerr effect or the photons with the Sagnac effect.} We find that the mean magnon number can selectively exhibit nonreciprocal linear or nonlinear (bistable) behavior with the strength of the strong driving field on the cavity. Assisted by this driving field, the magnon-phonon coupling is greatly enhanced, leading to the nonreciprocal photon-phonon entanglement via the swapping interaction between the magnons and photons. This nonreciprocal entanglement can be significantly enhanced with the magnon Kerr and Sagnac effects. Given the available parameters, the nonreciprocal photon-phonon entanglement can be preserved at $\sim3$ K, showing remarkable resilience against the bath temperature. The result reveals that our paper holds promise in developing various nonreciprocal devices with both the magnon Kerr and Sagnac effects in cavity magnomechanics.
  
	\end{abstract}
	
	\maketitle

	\section{Introduction}
	
Magnons~\cite{rameshti2022cavity,lachance2019hybrid,yuan2022quantum,prabhakar2009spin,van1958spin}, the quanta of collective spin excitations in magnetically ordered materials, especially for the yttrium iron garnet (YIG, $\rm {Y_3 Fe_5 O_{12}}$)~\cite{schmidt2020ultra,geller1957crystal,mallmann2013yttrium}, have drawn considerable attention theoretically and experimentally in quantum information science~\cite{rameshti2022cavity,li2020hybrid}. Thanks to high spin density and low collective loss~\cite{huebl2013high,tabuchi2014hybridizing,zhang2014strongly}, magnons in a YIG sphere can be strongly coupled to photons in microwave cavities for investigating various phenomena, such as dark modes~\cite{zhang2015magnon,bi2019magnon}, exceptional points~\cite{zhang2017observation,zhang2019experimental,zhang2019higher,zhao2020observation,cao2019exceptional,zhang2021exceptional,liu2019observation,sadovnikov2022exceptional,wang2023floquet}, nearly perfect absorption~\cite{rao2021interferometric}, unconventional magnon excitations~\cite{yuan2021unconventional}, stationary one-way quantum steering~\cite{yang2021controlling,guan2022cooperative}, and dissipative couplings~\cite{wang2019nonreciprocity,wang2020dissipative,harder2021coherent,yu2024nonhermitian}. With advanced experimental technologies, {the magnetostrictive force, originating from the deformation of the sphere's geometric structure during magnetization~\cite{Kittel1958}, gives rise to the nonlinear interaction between two modes (magnon mode and phonon mode), which was previously overlooked in commonly used dielectric or metallic materials and has been discovered in YIG spheres  recently~\cite{zhang2016cavity}}. This coupling mechanism allows magnons to interact with phonons in vibrated modes of the YIG sphere. Thus, a hybrid cavity magnomechanical (CMM)~\cite{zhang2016cavity} system is built when a YIG sphere meets a cavity. Obviously, this hybrid system combines the individual advantages of magnons, photons and phonons, providing great potential to investigate diverse quantum effects~\cite{li2018magnon,li2019squeezed,hatanaka2022chip,lu2021exceptional,kani2022intensive,huai2019enhanced,potts2022cavity}. Additionally, the magnon Kerr effect (i.e., the Kerr effect of magnons), {which denotes the nonlinear interaction among the magnon numbers} caused by the magnetocrystallographic anisotropy~\cite{zhang2019theory,prabhakar2009spin}, was experimentally demonstrated in the cavity magnonics frame~\cite{wang2018bistability,wang2016magnon}, giving rise to nonlinear cavity magnonics~\cite{zheng2023tutorial} as well as Kerr-modified CMM systems~\cite{shen2022mechanical}. This nonlinearity offers a great power in studying multistability~\cite{zhang2019theory,wang2018bistability,shen2022mechanical,bi2021tristability,yang2021bistability}, long-distance spin-spin interaction~\cite{xiong2022strong,xiong2023optomechanical,ji2023kerr,tian2023critical}, quantum phase transition~\cite{liu2023switchable,zhang2021parity}, and sensitive detection~\cite{zhang2023detection}.

In addition, macroscopic quantum entanglement has garnered significant attention in quantum information science\cite{PhysRevLett.109.013603,RevModPhys.81.865,lpor.200910010,RevModPhys.77.513,Kounalakisanalog,Zoubell,Zouquantum,Annaresolving}, owing to its wide applications in quantum transduction~\cite{tian2022experimental,zhong2022microwave}, quantum networking ~\cite{cirac1997quantum,kimble2008quantum,lodahl2017chiral}, quantum sensing ~\cite{degen2017quantum}, Bell-state tests~\cite{marinkovic2018optomechanical,vivoli2016proposal}, quantum teleportation ~\cite{hofer2015entanglement,hofer2011quantum}, and microwave-optics conversion~\cite{PhysRevA.93.023827,PhysRevA.96.013808,PhysRevA.106.032606}. To produce such entanglement, {\it nonlinear effects} are always required~\cite{Adesso_2007}. This results in the widespread exploration of macroscopic entanglement within nonlinear systems, including nonlinear cavity magnonics~\cite{zheng2023tutorial}, CMM systems~\cite{zuo2023cavity}, and cavity optomechanics (COM)~\cite{RevModPhys.86.1391}. Moreover, quantum entanglement can be well protected or enhanced in a spinning COM~\cite{PhysRevLett.125.143605}. {This is because the spinning COM allows the emergence of the strong correlation between the photon and the phonon in one chosen direction but weak correlation or complete lack of correlation in the opposite direction. Such direction-dependent entanglement is called nonreciprocal entanglement, which exploits the Sagnac-Fizeau effect to induce an opposite frequency shift on the cavity~\cite{Malykin2000The,Maayani2018Flying}. Specifically, when the rotation direction of the cavity field differs from the direction of the driving field, the light within the cavity undergoes varying equivalent refractive indices during propagation, resulting in an irreversible refractive index for the clockwise and counterclockwise modes. Correspondingly, the Lorentz reciprocity is broken and nonreciprocal entanglement emerges.}  Besides the Sagnac effect, the magnon Kerr effect can also be used to achieve nonreciprocal bipartite and tripartite entanglement in cavity optomechanics~\cite{PhysRevB.108.024105}; {that is, by only tuning the direction of the applied magnetic field along the crystallographic axis $[100]$ or $[110]$, asymmetric entanglement is produced.} This is due to the fact that the magnon Kerr effect can give rise to a positive or negative frequency shift on magnons as well as an additional parametric magnon amplifier under strong driving fields. However, nonreciprocal entanglement with these two nonlinear effects has yet to be revealed to date. 

Here we present a scheme to realize a nonreciprocal photon-phonon entanglement in a Kerr-modified spinning CMM system, where the magnon Kerr and the Sagnac effects are both considered. The proposed system consists of the Kerr magnons in the YIG sphere simultaneously coupled to photons in the spinning cavity via the magnetic dipole interaction and phonons in the mechanical mode via magnetostrictive interaction. In this system, the mean magnon number can selectively exhibit linear or nonlinear (bistable) nonreciprocal response under the strong driving field, where the nonreciprocity is induced by the Sagnac effect, and the linear (nonlinear) behavior is caused by the interplay between the magnon Kerr effect and the magnetostrictive interaction. With the enhanced magnetostrictive coupling between the magnons and phonons by the strong driving field, magnon mediated photon-phonon entanglement can be attained via the magnon-phonon and magnon-photon swapping interaction. This achieved entanglement can be nonreciprocally improved by both the magnon Kerr and Sagnac effects with varying the tunable system parameters. { To be more clear, we can interpret this nonreciprocal behavior from the view of our symmetrical operations on the proposed system. Without the magnon Kerr effect (or Sagnac effect), the symmetric photon-phonon entanglement is obtained when opposite magnetic fields along the crystallographic axis of the YIG sphere (or driving fields on the non-spinning cavity) are applied. However, the situation is changed when the magnon Kerr or Sagnac effect is taken into account, that is, the opposite magnetic fields (driving field) give rise to asymmetric entanglement. Since the generated entanglement is dependent on the direction of the magnetic field (driving field) applied to the YIG sphere (cavity), we here clarify that nonreciprocal entanglement is predicted in our proposed system.} Moreover, {such} nonreciprocal entanglement is robust against the bath temperature. With the available parameters, the photon-phonon entanglement can survive at $\sim3$ K, which is much higher than previous proposals. This paper provides opportunities for the development of diverse nonreciprocal devices in Kerr-modified spinning cavity magnomechanics.

This paper is organized as follows. In Sec.~\ref{sec2}, the model and the system Hamiltonian are described.  Then the steady-state solution and the effective Hamiltonian are given in Sec.~\ref{sec3}. In Sec.~\ref{sec4}, the nonreciprocal photon-phonon entanglement is studied with system parameters by taking both the Sagnac and magnon Kerr effects into account. Finally, a conclusion is given in Sec.~\ref{sec5}.

	\begin{figure}
		\includegraphics[scale=0.38]{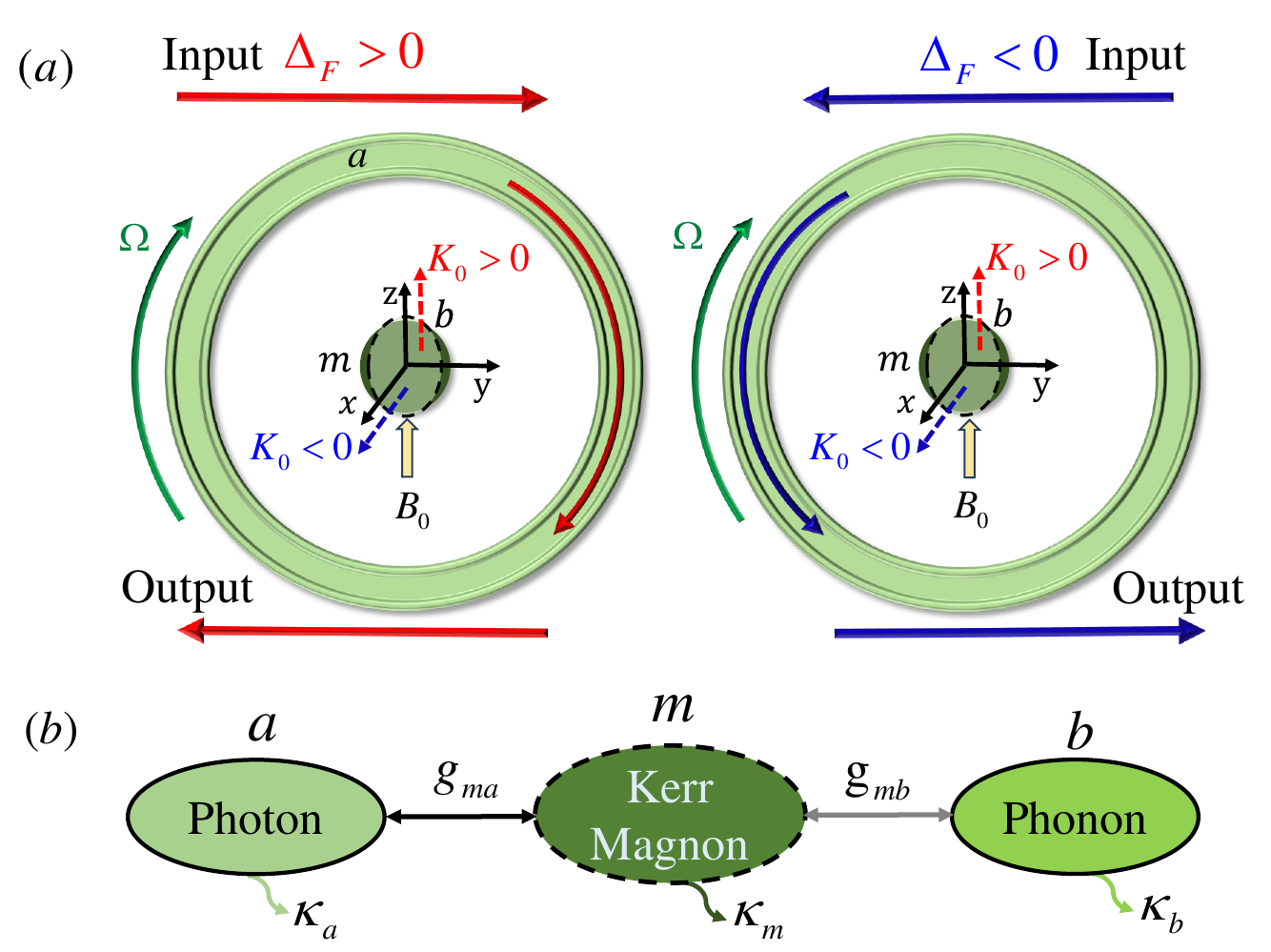}
		\caption{(a) Schematic of the Kerr-modified spinning CMM system. $K_0$ is the Kerr coefficient of the magnons, which can be tuned by the direction of the magnetic field $B_0$. When the magnetic field is aligned along the crystal axis $[100]~([110])$, $K_0>0~(<0)$.  For the spinning cavity, a positive (negative) frequency shift $\Delta_F$ is produced via the Sagnac effect when the driving field is clockwise (counterclockwise). (b) The coupling configuration. The Kerr magnons with the decay rate $\kappa_m$ are coupled to both the photons in the spinning cavity with the decay rate $\kappa_a$ and the phonons in the mechanical mode with the decay rate $\kappa_b$. The corresponding coupling strengths are $g_{ma}$ and $g_{mb}$.}\label{fig1}
	\end{figure}

	\section{The model and Hamiltonian}\label{sec2}

    We consider a hybrid Kerr-modified spinning cavity magnomechanical system consisting of a spinning resonator at an angular velocity $ \Omega$ holding photons coupled to Kerr magnons in the Kittel mode of a $\mu$m-YIG sphere, where the magnons of the YIG sphere placed in a static magnetic field $B_0$ are also coupled to phonons in the mechanical mode~[see Fig.~\ref{fig1}(a)].  The Hamiltonian of the proposed hybrid system can be written as (setting $\hbar=1$) 
    \begin{align}
    	H=H_{\rm SCM}+K_0 (m^\dag m)^2+i\epsilon_d(a^\dag e^{-i\omega_d t}-a e^{i\omega_d t}), \label{eq1}
    \end{align}
    with
    \begin{align}
    	H_{\rm SCM}=&(\omega_a-\Delta_F)a^\dag a+\omega_b b^\dag b+\omega_m m^\dag m\notag\\
    	&+g_{ma}(a^\dag m+a m^\dag)+g_{mb}m^\dagger m(b+b^\dagger),\label{eq2}
    \end{align}
    where $\omega_a$ is the resonance frequency of the non-spinning magnomechanical cavity, $\omega_b$ is the resonance frequency of the mechanical mode, and $\omega_m=\gamma H$ is the resonance frequency of the Kittel mode when the mechanical mode is at its equilibrium position, determined by the gyromagnetic ratio $\gamma$ and the external bias magnetic field $H$. $g_{ma}$ describes the coupling between the Kittel mode and the spinning cavity via the magnetic-dipole interaction, and $g_{mb}$ characterizes the single-magnon magnomechanical coupling between the Kittel mode and the mechanical mode via the magnetostrictive interaction~[see Fig.~\ref{fig1}(b)]. Experimentally, the strong magnon-photon coupling strength ($g_{ma}$) has been demonstrated~\cite{huebl2013high,tabuchi2014hybridizing,zhang2014strongly}, that is, $g_{ma}$ is larger than the dissipation rates of the cavity and Kittel modes, $\kappa_a$ and $\kappa_m$, i.e., $g_{ma}>\kappa_a,~\kappa_m$. Typically,  the magnon-phonon coupling $g_{mb}$ in the single-magnon level is weak, but it can be indirectly (directly) enhanced by imposing a strong driving field on the cavity (Kittel mode). The parameter $\Delta_F$ is the Sagnac-Fizeau shift of the cavity resonance frequency, induced by the light circulating in the spinning cavity, which can be given by~\cite{Malykin2000The,Maayani2018Flying} 
\begin{align}
\Delta_F=\pm\Omega\dfrac{nr\omega_a}{c}(1-\dfrac{1}{n^{2}}-\dfrac{\lambda}{n}\dfrac{d n}{d \lambda})\label{eq3}.
\end{align}
Here, $ n $ is the refractive index, $ r $ is the radius of the resonator, and $\lambda$ ($ c $) is the wavelength (speed) of the light in vacuum. The dispersion term $ dn/d\lambda $ in Eq.~(\ref{eq3}) denotes the relativistic origin of the Sagnac effect, which is small ($\sim1\%$)~\cite{Malykin2000The,Maayani2018Flying} and thus can be ignored. The sign $"+"~("-")$ in Eq.~(\ref{eq3}) corresponds to the clockwise (counterclockwise) driving field, where the direction of the cavity spinning is assumed to be along the clockwise direction. This means  $\Delta_F>0$ ( $\Delta_F<0$ ) for the clockwise (counterclockwise) driving field~[see Fig.\ref{fig1}(a)].

The second term in Eq.~(\ref{eq1}) related to $K_0$ depicts the Kerr nonlinearity of the magnons in the Kittel mode of the YIG sphere, arising from the magnetocrystallographic
anisotropy. The Kerr coefficient $K_0$ is inversely proportional to the volume of the YIG sphere~\cite{zhang2019theory}, and it can be tuned either positive or negative
by varying the direction of the static magnetic field~\cite{wang2018bistability}. Specifically, when the magnetic field is aligned along the crystallographic axis $[100]$ ($[110]$), we have $K_0>0~(<0)$~\cite{wang2018bistability}. Experimentally, $K_0$ can be tuned from $0.05$ to $100$ nHz for the diameter of the YIG sphere from $1$ mm to $100$ ${\rm \mu}$m. The last term in Eq.~(\ref{eq1}) is the Hamiltonian of the driving field acting on the spinning cavity, where $ \epsilon_d=\sqrt{2\kappa_a P/\omega_d}$ is the Rabi frequency, with $P$ being the power and $\omega_d$ the frequency. The operators $a~(a^\dag)$, $b~(b^\dag)$, and $m~(m^\dag)$ are the annihilation (creation) operators of the spinning cavity, the mechanical mode, and the Kittel mode, respectively. In the rotating frame with respect to $\omega_d$, the Hamiltonian in Eq.~(\ref{eq1}) becomes
\begin{align}
	\mathcal{H}=\mathcal{H}_{\rm SCM}+K_0 (m^\dag m)^2+i\epsilon_d(a^\dag-a),\label{eq4}
\end{align}
where $\mathcal{H}_{\rm SCM}=H_{\rm SCM}-\omega_d (a^\dag a+m^\dag m)$. 

\section{Quantum Langevin equation and the effective Hamiltonian}\label{sec3}

\subsection{Steady state solution}

By defining the frequency detuning of the spinning cavity (Kittel) mode from the driving field, i.e., $\Delta_{a(m)}=\omega_{a(m)}-\omega_d$, the dynamics of the considered system with dissipation can be governed by the quantum Langevin equations~\cite{1998Quantum}:
\begin{align}\label{eq5}
\dot{a}=&-[\kappa_a+i(\Delta_a-\Delta_F)]a-ig_{ma}m+\epsilon_d+\sqrt{2\kappa_a}a_{\rm in},\nonumber \\ 
\dot{b}=&-(\kappa_b+i\omega_b)b-ig_{mb}m^\dagger m+\sqrt{2\kappa_b}b_{\rm in}, \\
\dot{m}=&-(\kappa_m+i\Delta_m)m-ig_{ma}a-ig_{mb}m(b+b^\dagger)\nonumber \\
&-2iK_0m^\dagger mm+\sqrt{2\kappa_m}m_{\rm in}.\nonumber
\end{align}
Here $\sigma_{\rm in}~(\sigma=a,b,m)$ are the vacuum input noise operators of the spinning cavity, the mechanical mode, and the Kittel mode, respectively.  All these operators have zero mean values, i.e., $\langle \sigma_{\rm in}\rangle=0$. The correlation functions of these operators within the Markovian approximation satisfy
\begin{align}\label{eq6}
	\langle \sigma_{\rm in}^{ \dagger}(t^\prime)\sigma_{\rm in}(t)\rangle=&N_{\sigma}\delta(t-t^\prime),\notag\\
	\langle \sigma_{\rm in}(t)\sigma_{\rm in}^{\dagger}(t^\prime)\rangle=&(N_{\sigma}+1)\delta(t-t^\prime),
\end{align} 
where $ N_\sigma=[{\rm exp}({\hbar\omega_\sigma}/{k_{B}T}-1)]^{-1}$ is the mean thermal excitation number in the mode $\sigma$, with $k_B$ being
the Boltzmann constant and $T$ the bath temperature.

By rewriting each operator ($\sigma$) as the sum of its expectation ($\sigma_s$) and fluctuation ($\delta \sigma$) in Eqs.~(\ref{eq5}), i.e., { $\sigma\rightarrow\sigma_s+\delta\sigma$}, a set of equations related to the operator expectation can be given by
\begin{align}\label{eq7}
	\dot{a}_s=&-[\kappa_a+i(\Delta_a-\Delta_F)]a_s-ig_{ma}m_s+\epsilon_d,\nonumber \\ 
	\dot{b}_s=&-(\kappa_b+i\omega_b)b_s-ig_{mb}|m_s|^2, \\
	\dot{m}_s=&-[\kappa_m+i(\tilde{\Delta}_m+\Delta_K)]m_s-ig_{ma}a_s,\nonumber
\end{align}
where $\tilde{\Delta}_m=\Delta_m+2g_{mb}{\rm Re}[b_s]$ is the frequency detuning induced by the displacement of the mechanical mode ($2g_{mb}{\rm Re}[b_s]$), and $\Delta_K=2K_0|m_s|^2$ is the frequency shift caused by the magnon Kerr effect. In the long-time limit, the proposed system reaches its steady state, i.e., $\dot{\sigma}_s=0$, so Eqs.~(\ref{eq7}) reduce to
\begin{align}\label{eq8}
		[\kappa_a+i(\Delta_a-\Delta_F)]a_s+ig_{ma}m_s-\epsilon_d&=0,\nonumber \\ 
		(\kappa_b+i\omega_b)b_s+ig_{mb}|m_s|^2&=0, \\
		[\kappa_m+i(\tilde{\Delta}_m+\Delta_K)]m_s+ig_{ma}a_s&=0.\nonumber
\end{align}
By directly solving these equations, we have
\begin{align}\label{eq9}
a_s=&\frac{\epsilon_d-ig_{ma}m_s}{\kappa_a+i(\Delta_a-\Delta_F)},\notag\\
b_s=&-\frac{ig_{mb}\abs{m_s}^2}{\kappa_b+i\omega_b},\\
m_s=&-\frac{ig_{ma}a_s}{\kappa_m+i(\tilde{\Delta}_m+\Delta_K)}.\notag
\end{align}
Since $\Delta_F>0~(<0)$ is dependent on the direction of the clockwise (counterclockwise) driving field, so the mean photon number ($|a_s|^2$) has different values for the opposite driving fields, indicating that $|a_s|^2$ in the spinning cavity behaves nonreciprocally. This nonreciprocity can indirectly give rise to nonreciprocal mean magnon ($|m_s|^2$) and phonon ($|b_s|^2$) numbers because of the direct coupling between the spinning cavity and the Kittel mode of the YIG sphere, and the magnon-mediated coupling between the spinning cavity and the mechanical mode [see the last two equations in Eqs.~(\ref{eq9})]. It is worth mentioning that such a nonreciprocal situation can also be induced by the magnon Kerr effect, even for a non-spinning cavity ($\Delta_F=0$). This is because $K_0>0$ or $K_0<0$, depending on the direction of the magnetic field, leads to $\Delta_K>0$ or $\Delta_K<0$. Thus, nonreciprocal mean magnon number is directly obtained [see the last equation in Eqs.~(\ref{eq9})] and causes the nonreciprocal mean photon and phonon numbers.

\subsection{Nonreciprocal bistability}

From Eqs.~(\ref{eq9}),  a cubic equation related to the mean magnon number $|m_s|^2=M$ can be given by
\begin{align}
 [\kappa_m^{\prime2}+(\Delta_m^\prime+K_0^\prime M)^2]M=\eta_a \epsilon_d^2,\label{eq10}
\end{align}
where
\begin{align}\label{eq11}
	\kappa_m^{\prime}&=\kappa_m+\eta_a\kappa_a,\notag\\
	\Delta_m^\prime&=\Delta_m-\eta_a(\Delta_a-\Delta_F),\notag\\
	K_0^\prime&=2(K_0-\eta_b\omega_b),\\
	\eta_a&=\frac{g_{ma}^2}{\kappa_a^2+(\Delta_a-\Delta_F)^2},\notag\\
	\eta_b&=\frac{g_{mb}^2}{\kappa_b^2+\omega_b^2}.\notag
\end{align}
\begin{figure}
	\includegraphics[scale=0.35]{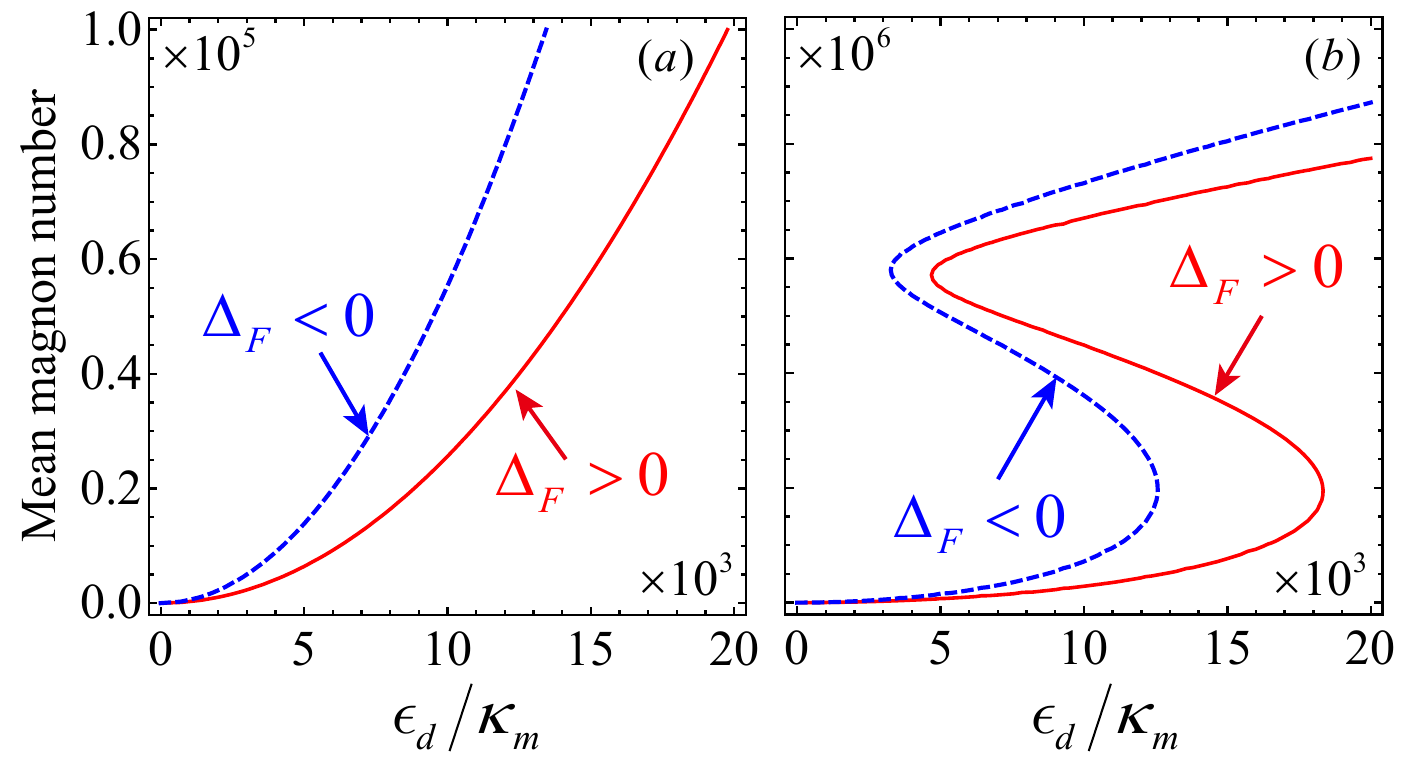}
	\caption{The mean magnon number vs the normalized amplitude of the driving field with (a) $K_0=\eta_b\omega_b$ and (b) $K_0=0.1\eta_b\omega_b$, where $\omega_b/2\pi=10$ MHz, $ \kappa_a=\kappa_m=0.1\omega_b$, $\kappa_b /2\pi=100$ Hz, $ g_{\rm ma}=0.2\omega_b$, $g_{mb}=10^{-3}\omega_b$, $|\Delta_F|=0.2\omega_b$, $\Delta_a=-\omega_b$, and $\Delta_m=\omega_b$. } \label{fig2}
\end{figure}
As $K_0$ can be tuned via adjusting the direction of the magnetic field, so $K_0^\prime$ in Eq.~(\ref{eq10}) can be zero or nonzero. This is because pure magnomechanical coupling [$m^\dag m(b+b^\dag)$], similar to optomechanics, is equivalent to an effective magnon Kerr Hamiltonian by performing the unitary transformation $U=\exp[m^\dag m(b-b^\dag)]$~\cite{2008Effective}. When $K_0^\prime=0$, the cubic equation given by Eq.~(\ref{eq10}) becomes a linear equation in the mean magnon number $M$. Obviously, it is proportional to the square of the Rabi frequency of the driving field and nonreciprocally responses to the driving fields from opposite directions, as shown in Fig.~\ref{fig2}(a). When $K_0^\prime\neq0$, the mean magnon number $M$ determined by Eq.~(\ref{eq10}) can have two switching points for bistability under  specific parameter conditions [see Fig.~\ref{fig2}(b)], at which there must be $d\epsilon_d/dM=0$, i.e.,
\begin{align}\label{eq12}
	3K_0^{\prime2} M^2+4K_0^\prime\Delta_m^\prime M+\kappa_m^{\prime2}+\Delta_m^{\prime2}=0.
\end{align}
This equation has two real roots, corresponding to two switching points of the bistability, only when the root discriminant satisfies the inequality $\Delta_m^{\prime2}-3\kappa_m^{\prime2}>0$, i.e.,
\begin{align}
	\Delta_m^{\prime}>\sqrt{3}\kappa_m^{\prime}~{\rm or}~\Delta_m^{\prime}<-\sqrt{3}\kappa_m^{\prime}.
\end{align}
In particular, when $\Delta_m^{\prime2}-3\kappa_m^{\prime2}=0$, i.e., $\Delta_m^{\prime}=\pm\sqrt{3}\kappa_m^{\prime}$, Eq.~(\ref{eq12}) has two equal real roots ($M_{1(2)}=-2\Delta_m^\prime/3K_0^\prime$), that is, two switching points coalesce to one point, indicating no bistability. This give rise to a critical driving strength,
\begin{align}\label{eq14}
	\epsilon_d^c=\sqrt{-8\kappa_m^{\prime2}\Delta_m^\prime/9\eta_aK_0^\prime}=\sqrt{\pm8\kappa_m^{\prime3}/9\eta_aK_0^\prime},
\end{align}
where the positive (negative) symbol denotes $K_0^\prime>0$ and $\Delta_m^\prime<0$ ($K_0^\prime<0$ and $\Delta_m^\prime>0$). Eq.~(\ref{eq14}) indicates that magnonic bistability can be predicted when the strength of the driving field exceeds the critical value, i.e., $\epsilon_d>\epsilon_d^c$, as shown in Fig.~\ref{fig2}(b). Due to different responses of the spinning cavity to the CW or CCW driving field, nonreciprocal bistability can be apparently observed [see the red and blue curves in Fig.~\ref{fig2}(b)].

\subsection{Fluctuation dynamics}
Apart from the steady-state dynamics when the transformation { $\sigma\rightarrow\sigma_s+\delta\sigma$} is substituted into Eqs.~(\ref{eq5}), the fluctuation dynamics can also be obtained:
{\begin{align}\label{eq15}
	\delta\dot{a}=&-[\kappa_a+i(\Delta_a-\Delta_F)] \delta a-ig_{ma}  \delta m+\sqrt{2\kappa_a}a_{\rm in},\nonumber \\ 
	\delta\dot{b}=&-(\kappa_b+i\omega_b)\delta  b-i(G_{mb} \delta m^\dag+G_{mb}^*\delta  m)+g_{mb} \delta m^\dag \delta m\notag\\
	&+\sqrt{2\kappa_b}b_{\rm in}, \\
	\delta\dot{m}=&-[\kappa_m+i(\tilde{\Delta}_m+2\Delta_K)] \delta m-ig_{ma} \delta a+\sqrt{2\kappa_m}m_{\rm in}\notag\\
	&-ig_{mb} \delta m( \delta b +\delta  b^\dagger)-2iK_0 m_s^2 \delta m^\dag-iG_{mb}( \delta b+ \delta b^\dagger)\nonumber \\
	&-2iK_0(m_s^* \delta m^2+2m_s \delta m^\dag  \delta m+ \delta m^\dag  \delta m^2),\nonumber
\end{align}}
where $G_{mb}=g_{mb}m_s$ is the effective magnomechanical coupling strength significantly enhanced by multiple magnons. Below we assume that $ m_s $ is real for simplicity. This can be realized by choosing the {proper} phase of the driving field according to Eqs.~(\ref{eq9}). Under the strong driving field, {the condition $|a_s|\gg1$ can be realized. Because of the beam-splitter interaction ($\propto a^\dag m+a m^\dag$) between the photons and magnons, we have $|m_s|\gg 1$, which directly requires the system parameters to satisfy
\begin{align}\label{eq}
	\varepsilon_d^2g_{ma}^2\gg\abs{[\kappa_m+i(\tilde{\Delta}_m+\Delta_k)][\kappa_a+i(\Delta_a-\Delta_F)]+g_{ma}^2}^2
\end{align}
according to Eqs.~(\ref{eq9}). This indicates that the mean-field approximation can be applied to Eqs.~(\ref{eq15}) when Eq.~(\ref{eq}) is kept, so that the high-order fluctuations in Eq.~(\ref{eq15}) can be safely ignored}. As a result,  Eqs.~(\ref{eq15}) reduces to
{\begin{align}\label{eq16}
	\delta \dot{a}=&-[\kappa_a+i(\Delta_a-\Delta_F)] \delta a-ig_{ma}  \delta m+\sqrt{2\kappa_a}\delta a_{\rm in},\nonumber \\ 
	\delta \dot{b}=&-(\kappa_b+i\omega_b)\delta b-iG_{mb}( \delta m^\dag+ \delta m)+\sqrt{2\kappa_b}b_{\rm in},\notag \\
	\delta \dot{m}=&-[\kappa_m+i(\tilde{\Delta}_m+2\Delta_K)] \delta m-i\Delta_K \delta m^\dag-ig_{ma} \delta a\notag\\
	&-iG_{mb}( \delta b+ \delta b^\dagger)+\sqrt{2\kappa_m}m_{\rm in}.
\end{align}}
We then rewrite the above equations as { $ \delta \dot{\sigma}=i[H_{\rm eff},\delta \sigma]-\kappa_\sigma \delta \sigma+\sqrt{2\kappa_\sigma} \sigma_{\rm in}$}, so the effective Hamiltonian of the linearized system can be given by
{\begin{align}\label{eq17}
	H_{\rm eff}=&(\Delta_a-\Delta_F) \delta a^\dag \delta  a+\omega_b  \delta b^\dag b+(\tilde{\Delta}_m+2\Delta_K)  \delta m^\dag \delta  m\notag\\
	&+g_{ma}( \delta a^\dag  \delta m+ \delta a \delta m^\dag)+\frac{\Delta_K}{2}( \delta m^\dag \delta m^\dag+ \delta m\delta m)\notag\\
	&+G_{mb}( \delta m^\dag+\delta m)( \delta b^\dag+ \delta b).
\end{align}}
Note that the two-magnon effect (i.e., $ \delta m^\dag \delta m^\dag+\delta m \delta m$) stems from the magnon Kerr nonlinearity in the presence of the strong driving field, which can be well tuned by varying the strength of the driving field.

\section{The nonreciprocal photon-phonon entanglement} \label{sec4}
\subsection{{ Entanglement calculation}}

With the effective Hamiltonian $H_{\rm eff}$ in hand, its dynamics governed by Eqs.~(\ref{eq16}) can be rewritten in a more compact form as $ \dot{u}(t)=Au(t)+f(t) $, where
$ u^T(t)=(X_a, Y_a, X_m, Y_m, X_b, Y_b)$ is the vector operator of the system, $f^T(t)=(\sqrt{2\kappa_a}X_{a}^{\rm in}$, $\sqrt{2\kappa_a}Y_{a}^{\rm in}$, $\sqrt{2\kappa_m}X^{\rm in}_{m}$, $\sqrt{2\kappa_m}Y^{\rm in}_{m}$, $\sqrt{2\kappa_b}X^{\rm in}_{b}$, $\sqrt{2\kappa_b}Y^{\rm in}_{b})$ is the input noise of the system, and
\begin{equation}\label{eq18}
\small{
	\setlength{\arraycolsep}{0.01pt}
A=\left(\begin{array}{cccccc}
-\kappa_a  &{\Delta}_a-\Delta_F&0&g_{ma}& 0&  0\\
-{\Delta}_a+\Delta_F  &-\kappa_a&-g_{\rm ma} &0 &0&0\\
0  &g_{ma} &-\kappa_m&\tilde{\Delta}_m+\Delta_K &0 &0\\
-g_{ma}&0&-\widetilde{\Delta}_m-3\Delta_K&-\kappa_m&-2G_{mb} &0\\
0 &0&0 &  0  & -\kappa_b  &\omega_b\\
0 &0&-2 G_{mb}&0 & -\omega_b & -\kappa_{b}
	\end{array}\right)}
\end{equation}
is the drift matrix. Here { $ X_\sigma=(\delta \sigma^\dagger+\delta \sigma)/\sqrt{2} $, $ Y_\sigma=i(\delta \sigma^\dagger-\delta \sigma)/\sqrt{2} $,} $ X_{\sigma}^{\rm in}=(\sigma_{\rm in}^\dagger+\sigma_{\rm in})/\sqrt{2} $, and $ Y_\sigma^{\rm in}=i(\sigma_{\rm in}^{\dagger}-\sigma_{\rm in})/\sqrt{2} $.

Since the input quantum noises are zero-mean quantum Gaussian noises, the quantum steady state for the fluctuations is a zero-mean continuous variable Gaussian state, fully characterized
by a $ 6\times6 $ correlation matrix $V_{ij}^{(6)}=\langle u_{i}(\infty)u_{j}(\infty)+u_{j}(\infty)u_{i}(\infty)\rangle$ ($i,j=1,2,\ldots,6$). The matrix $V$ can be  obtained by directly solving the Lyapunov equation,
\begin{align}\label{eq19}
AV+VA^\dagger=-D,
\end{align}
 where the diffusion matrix $ D={\rm diag}[ \kappa_a(2 N_a+1), \kappa_a(2N_a+1), \kappa_m(2N_m+1), \kappa_m(2N_m+1), \kappa_b(2N_b+1), \kappa_{b}(2N_b+1)] $ is defined by $ D_{ij}  \delta(t-t^\prime)=\langle v_{i}(t)v_j(t^\prime)+v_j(t^\prime)v_{i}(t)\rangle/2 $. Once the matrix $V$ is obtained, one can investigate arbitrary bipartite entanglement of interest in the proposed system via the logarithmic negativity
 \begin{align}\label{eq20}
 	E_N\equiv \rm max[0,-ln2\eta^-],
 \end{align}
with
\begin{align}\label{eq21}
	\eta^-=2^{-1/2}[\Sigma-(\Sigma^2-4\rm det {V}_4)^{1/2}]^{1/2},
\end{align}
where $ \Sigma=\rm{det} \mathcal{A}+\rm{det} \mathcal B-2\rm{det} \mathcal C $ and 
${V}_4=\begin{pmatrix}
\mathcal A &\mathcal C\\
\mathcal {C^T}& \mathcal B\\
\end{pmatrix}
$ is the $ 4 \times4 $ block form of the correlation matrix, associated with two modes of interest. $\mathcal{A}$, $\mathcal{B}$, and $\mathcal{C}$ are the $ 2\times2 $ blocks of $V_4$. A positive logarithmic negativity ($E_N>0$) denotes the presence of bipartite entanglement of the interested two modes in the considered system.

\subsection{{Nonreciprocal entanglement exploration}}

\begin{figure}
	\includegraphics[scale=0.44]{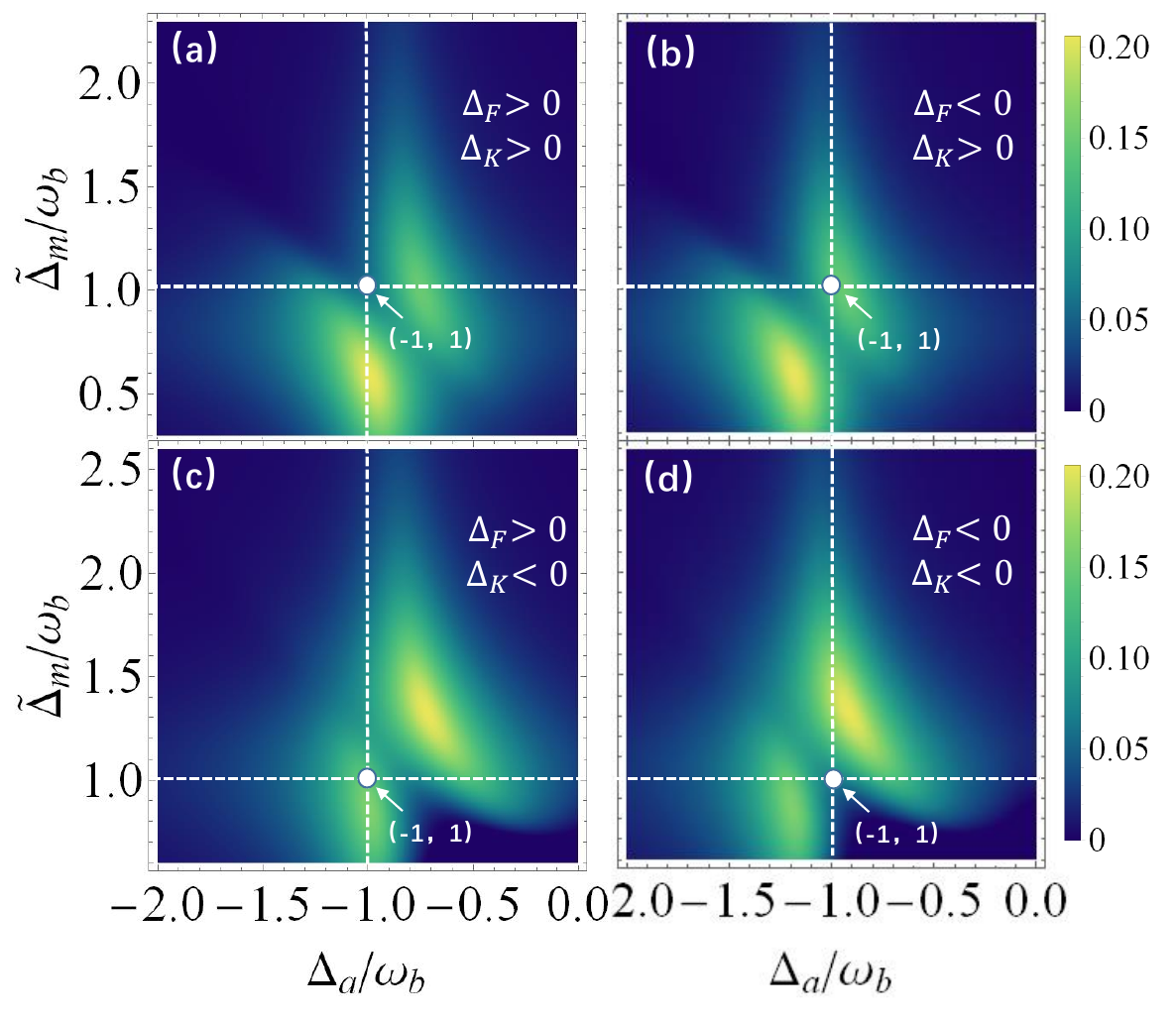}
	\caption{Density plot of the photon-phonon entanglement $ E_{ab} $ as functions of $\Delta_a/\omega_b$ and $\tilde{\Delta}_m/\omega_b$ with (a) $\Delta_K>0$, $\Delta_F>0$, (b) $\Delta_K>0$, $\Delta_F<0$, (c) $\Delta_K<0$, $\Delta_F>0$, and (d) $\Delta_K<0$, $\Delta_F<0$. Other parameters are the same as those in Fig.~\ref{fig2} except for $G_{mb}=0.2\omega_b$, $|\Delta_F|=|\Delta_K|=0.1\omega_b$, and the bath temperature $T=10$ mK. 
	} \label{fig3}
\end{figure}

We first plot the logarithmic negativity $E_{ab}$ as functions of the normalized $\Delta_a/\omega_b$ and $\tilde{\Delta}_m/\omega_b$ in the presence of both the Sagnac and the magnon Kerr effects in Fig.~\ref{fig3}. The chosen parameters are the same as those in Fig.~\ref{fig2} except for $G_{ mb}=0.2\omega_b$, $|\Delta_F|=|\Delta_K|=0.1\omega_b$ and the bath temperature $ T=10$ mK. These parameters can ensure the system is stable according to the Routh-Hurwitz criterion. 

From Fig.~\ref{fig3}, one can see that the photon-phonon entanglement can be tuned by changing the frequency detunings $\Delta_a$ and $\tilde{\Delta}_m$. In particular, its optimal value is predicted at $\Delta_a\approx-\omega_b+\Delta_F$ and $\tilde{\Delta}_m\approx\omega_b-2\Delta_K$. The mechanism of this optimal entanglement can be interpreted as follows: When $\tilde{\Delta}_m\approx\omega_b-2\Delta_K$, the magnomechanical subsystem is driven to the red sideband, where the mechanical mode can be well cooled for allowing considerable magnon-phonon entanglement owing to the enhanced magnomechanical coupling $G_{mb}$. Then the magnon-phonon entanglement is significantly transferred to the photon-phonon entanglement via the beam-splitter magnon-photon interaction at $\Delta_a\approx-\omega_b+\Delta_F$. Moreover, from Fig.~\ref{fig3}, we also find that the predicted photon-phonon entanglement nonreciprocally responds to the change of the frequency detunings $\Delta_K$ or $\Delta_F$, i.e., the magnon Kerr or the Sagnac effects. This means that the nonreciprocal photon-photon entanglement can be achieved by including these two effects. Specifically, when $\Delta_F>0$, the optimal cavity frequency detuning is fixed at $\Delta_a\approx-0.8\omega_b$, but the optimal magnon frequency detuning is $\tilde{\Delta}_m\approx0.6\omega_b$ for $\Delta_K>0$ [Fig.~\ref{fig3}(a)] and $\tilde{\Delta}_m\approx1.4\omega_b$ for $\Delta_K<0$ [Fig.~\ref{fig3}(c)]. This indicates that the photon-phonon entanglement nonreciprocally changes with the magnon Kerr effect when the Sagnac effect is fixed. A similar result can also be obtained from Figs.~\ref{fig3}(b) and \ref{fig3}(d). By comparing Figs.~\ref{fig3}(a) with \ref{fig3}(b)~[or Figs.~\ref{fig3}(c) with \ref{fig3}(d)], the optimal photon-phonon entanglement has left shifts on the frequency detuning $\Delta_a$, which means the nonreciprocal photon-phonon entanglement can be induced by the Sagnac effect when the magnon Kerr effect is fixed. 

\begin{figure}
	\includegraphics[scale=0.34]{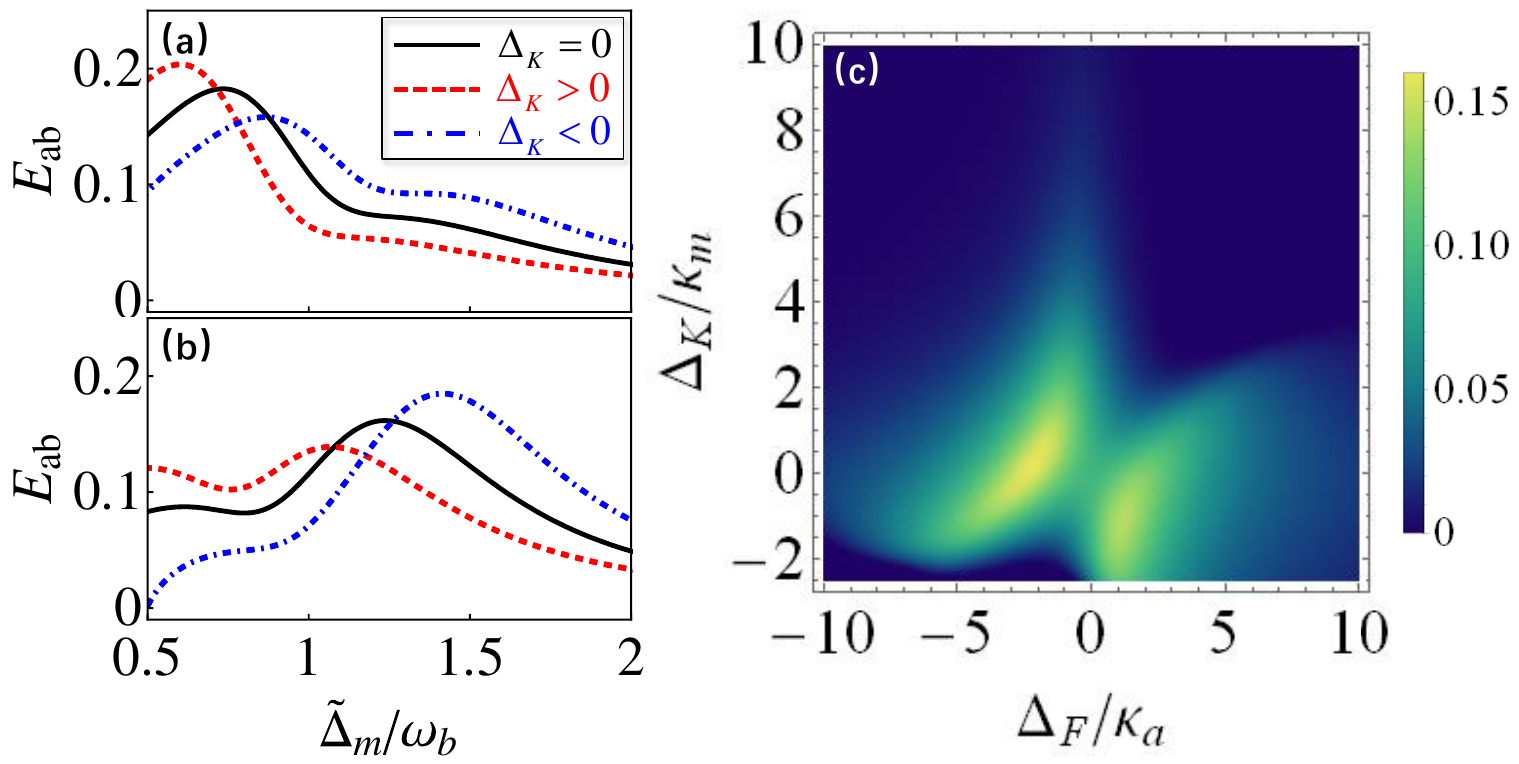}
	\caption{The photon-phonon entanglement vs $\tilde{\Delta}_m/\omega_b$ with and without the magnon Kerr effect (a, b) and $\Delta_K$ and $\Delta_F$ (c), where (a) $\Delta_F>0$ and (b) $\Delta_F<0$. Other parameters in panels (a)-(c) are the same as those in Fig.~\ref{fig3}.}
	\label{fig4}
\end{figure}
Figures~\ref{fig4}(a) and \ref{fig4}(b) further show the nonreciprocal behavior of the photon-phonon entanglement with the normalized  $\tilde{\Delta}_m/\omega_b$ with both the nonlinear effects, where $\Delta_a=-\omega_b$ is fixed. For $\Delta_F>0$ [see Fig.~\ref{fig4}(a)], we find that the photon-phonon entanglement can be nonreciprocally enhanced ($\Delta_K>0$) or reduced ($\Delta_K>0$), compared to the case without the magnon Kerr effect. In the case of $\Delta_F<0$ [see Fig.~\ref{fig4}(b)], the situation becomes opposite. From Fig.~\ref{fig4}(a) and \ref{fig4}(b), one can see that the photon-phonon entanglement can also be nonreciprocally enhanced or reduced by the Sagnac effect with the magnon Kerr effect.  Figure~\ref{fig4}(c) directly shows the behavior of the photon-phonon entanglement with these two effects. We show that the nonreciprocal photon-phonon entanglement can be predicted in a broad range of $\Delta_K$ and $\Delta_F$, and its optimal value can be obtained around the region of $\Delta_K\Delta_F<0$.

\subsection{{The magnomechanical coupling effect}}
\begin{figure}
	\includegraphics[scale=0.36]{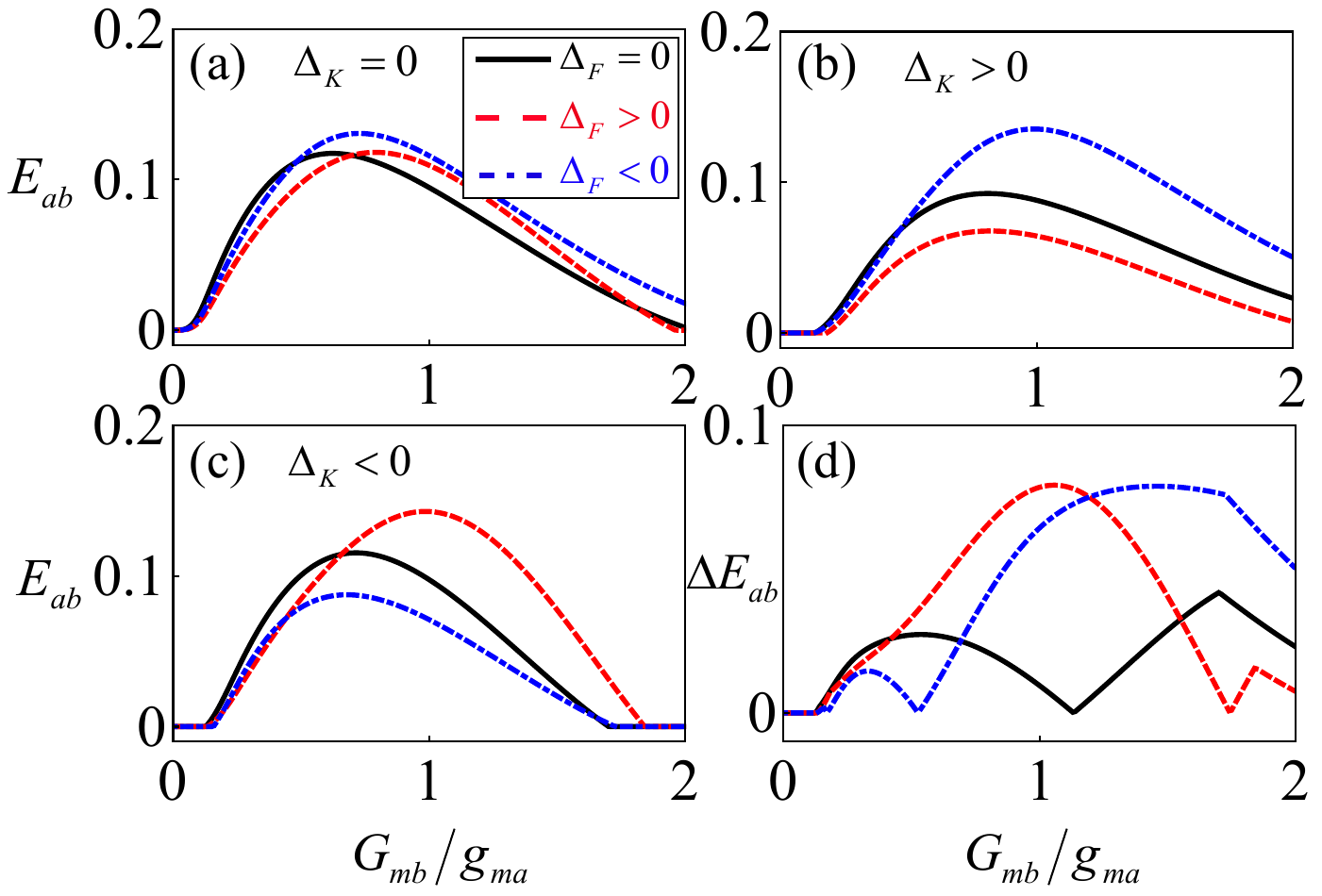}
	\caption{The photon-phonon entanglement vs $G_{mb}/g_{ma}$ with (a) $\Delta_K=0$, (b) $\Delta_K>0$, and (c) $\Delta_K<0$, {where} the Sagnac effect is considered. (d) The nonreciprocity of the photon-phonon entanglement induced by the magnon Kerr effect vs $G_{mb}/g_{ma}$ with and without the Sagnac effect. Other parameters in panels (a)-(c) are the same as those {in Fig.~\ref{fig3}.}} \label{fig5}
\end{figure}

In fact, the magnomechanical coupling strength $G_{mb}$ can be fine tuned by adjusting the amplitude of the driving field on the cavity [see Eqs.~(\ref{eq9})] in our proposal. So how does the magnomechanical coupling affect the photon-phonon entanglement with or without the magnon Kerr and Sagnac effects? To show this, we plot $E_{ab}$ versus the normalized coupling strength $G_{mb}/g_{ma}$ in Figs.~\ref{fig5}(a-c). Obviously, the photon-phonon entanglement increases first to its maximal value and then decreases to zero with the magnomechanical coupling strength $G_{mb}/g_{ma}$. Specifically, the Sagnac effect can only give a weak nonreciprocity on the photon-phonon entanglement in the absence of the magnon Kerr effect $\Delta_K=0$ [see Fig.~\ref{fig5}(a)]. In the presence of the magnon Kerr effect ($\Delta_K\neq0$), we find that a visible nonreciprocity on the photon-phonon entanglement can be induced by the Sagnac effect [see Fig.~\ref{fig5}(b) or \ref{fig5}(c)]. This is because the photon-phonon entanglement can be significantly enhanced (reduced)  when $\Delta_K\Delta_F<0$ ($\Delta_K\Delta_F>0$). From Figs.~\ref{fig5}(a-c), the magnon Kerr effect induced nonreciprocity on the photon-phonon entanglement can also be revealed with or without the Sagnac effect. More intuitively, we plot the difference ($\Delta E_{ab}$) of the logarithmic negativities between the cases of $\Delta_K>0$ and $\Delta_K<0$ in Fig.~\ref{fig5}(d), where $\Delta E_{ab}$ is defined as
\begin{align}
	\Delta E_{ab}=|E_{ab}(\Delta_K>0)-E_{ab}(\Delta_K<0)|.\label{eq22}
\end{align}
When the Sagnac effect is included, we find that large $\Delta E_{ab}$ can be obtained, as shown by the red square and blue diamond curves.

\begin{figure}
	\includegraphics[scale=0.36]{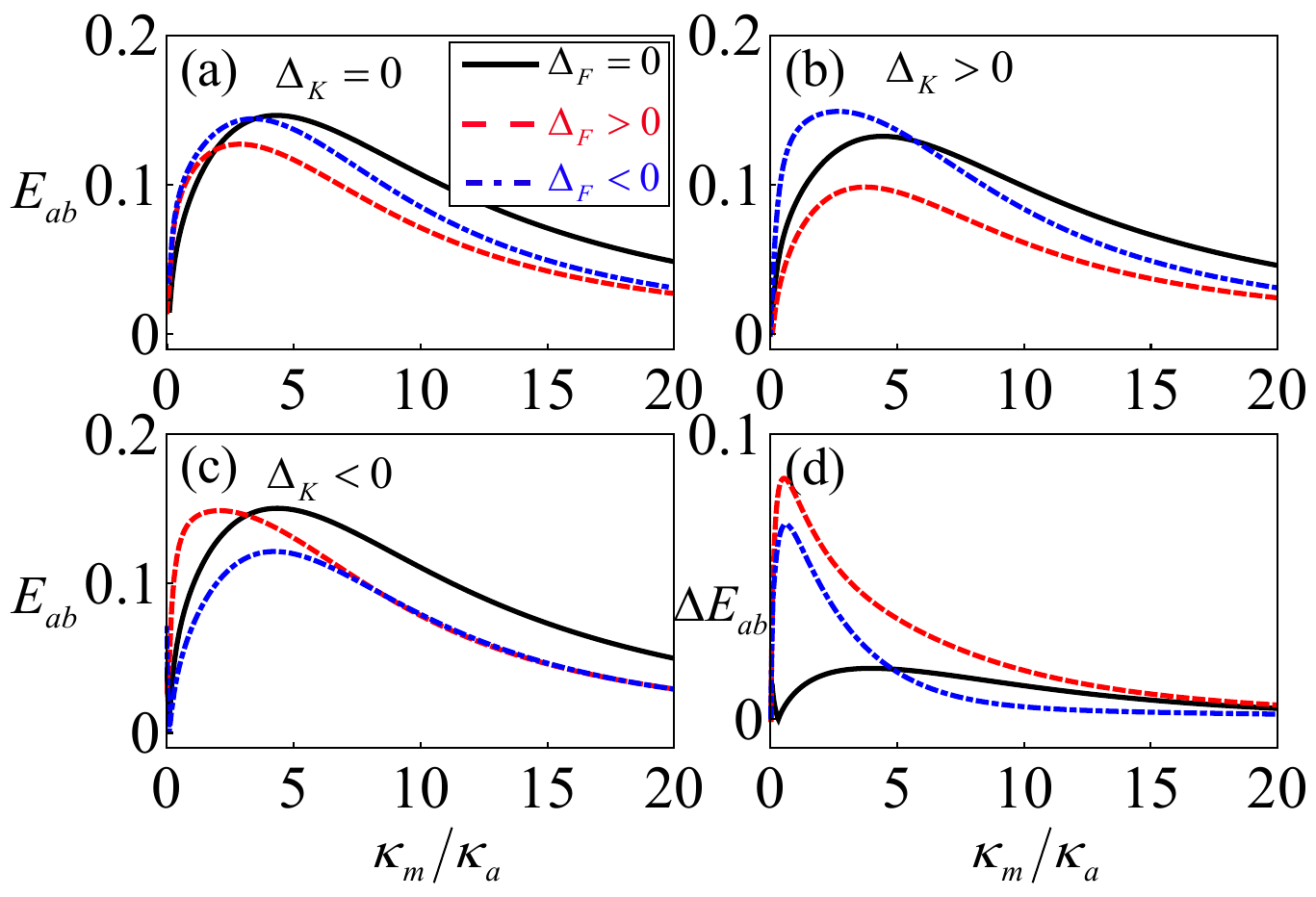}
	\caption{The photon-phonon entanglement vs $\kappa_m/\kappa_a$ with (a) $\Delta_K=0$, (b) $\Delta_K>0$, and (c) $\Delta_K<0$, where the Sagnac effect is considered. (d) The nonreciprocity of the photon-phonon entanglement induced by the magnon Kerr effect vs $\kappa_m/\kappa_a$ with and without the Sagnac effect. Other parameters in panels (a)-(c) are the same as those in Fig.~\ref{fig3}.}\label{fig6}
\end{figure}

\subsection{{The magnon decay rate effect}}

Besides the magnomechanical coupling strength, the decay rate of the magnons in the Kittel mode of the YIG sphere can also be adjusted experimentally via changing the distance between the YIG sphere and the microwave antenna. We find that the optimal photon-phonon entanglement can be realized by tuning $\kappa_m$ when other parameters are fixed, as shown in Figs.~\ref{fig6}(a-c). In the absence of the magnon Kerr effect, i.e., $\Delta_K=0$, one can see that the photon-phonon entanglement is robust against the Sagnac effect for the small decay rate of the Kittel mode [see Fig.~\ref{fig6}(a)], but when the magnon Kerr effect is taken into account, i.e., $\Delta_K\neq0$, the visible nonreciprocity induced by the Sagnac effect can be predicted [see Fig.~\ref{fig6}(b) or \ref{fig6}(c)]. We also show that the nonreciprocity induced by the Sagnac effect in the presence or absence of the magnon Kerr effect can be suppressed by increasing $\kappa_m$. This means that the nonreciprocity induced by the Sagnac effect can only be observed for the proper value of $\kappa_m$.
A similar result can also be obtained for the magnon Kerr effect induced nonreciprocity of the photon-phonon entanglement with or without the Sagnac effect [see Fig.~\ref{fig6}(d)].

\subsection{{The temperature effect}}

\begin{figure}
	\includegraphics[scale=0.36]{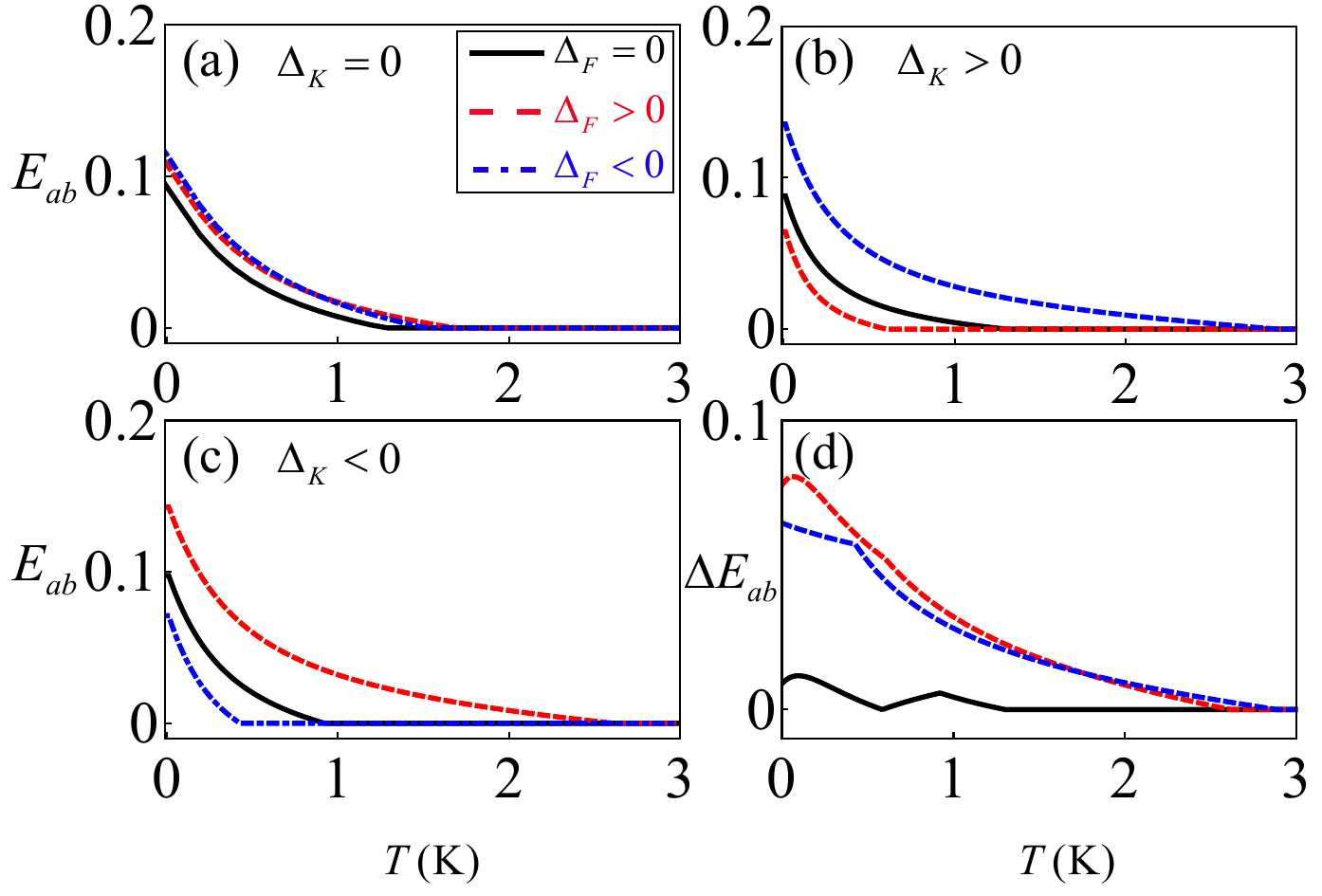}
	\caption{The photon-phonon entanglement vs the bath temperature $T$ with (a) $\Delta_K=0$, (b) $\Delta_K>0$, and (c) $\Delta_K<0$, {where} the Sagnac effect is considered. (d) The nonreciprocity of the photon-phonon entanglement induced by the magnon Kerr effect versus $T$ with and without the Sagnac effect. Other parameters in panels (a)-(c) are the same as those in Fig.~\ref{fig3}.} \label{fig7}
\end{figure}

Next, we check the effect of the bath temperature on the photon-phonon entanglement in our proposal. For this, we plot the logarithmic negativity $E_{ab}$ versus the temperature $T$ with or without the magnon Kerr or the Sagnac effect in Figs.~\ref{fig7}(a-c).  Fig.~\ref{fig7}(a) shows that the Sagnac effect can only cause a slight {improvement on both the photon-phonon entanglement and its survival temperature at $\Delta_K=0$. Meanwhile, we also observe the same slight improvement on both the photon-phonon entanglement and its survival temperature, as indicated by the black lines in Figs.~\ref{fig7}(a-c) at $\Delta_F=0$. But when both the magnon Kerr and Sagnac effects are included, we find that the photon-phonon entanglement and its survival temperature can be significantly improved (reduced) at $\Delta_K\Delta_F<0~(\Delta_K\Delta_F>0)$ [see the blue line in Figs.~\ref{fig7}(b) and the red line in Figs.~\ref{fig7}(c)]. This indicates that the high survival temperature for the photon-phonon entanglement can only be obtained by utilizing the negative synergistic effect (i.e., the coefficients of the Kerr and Sagnac effects have opposite signs). When the positive synergistic effect (i.e., the coefficients of the Kerr and Sagnac effects have the same signs) of the Kerr and Sagnac effects is considered, the entanglement and its survival temperature reduce.}  Notably, the survival temperature of the photon-phonon entanglement in our proposal can be improved to $\sim 3$ K, which is about $15$ times more than the previous proposal~\cite{li2018magnon}. Figure ~\ref{fig7}(d) also demonstrates that the magnon Kerr effect has the same effect on photon-phonon entanglement as the Sagnac effect.  

 \section{CONCLUSION}\label{sec5}
 
In summary, we have proposed to generate a nonreciprocal photon-phonon entanglement in a Kerr-modified spinning cavity magnomechanics. The mean magnon number here can selectively display nonreciprocal linear or nonlinear (bistable) behavior with the strength of the strong driving field, where the nonreciprocity arises from the Sagnac effect, and the linear (nonlinear) behavior is the result of the interplay between the magnon Kerr effect and the magnetostrictive effect. With the enhanced magnon-phonon coupling and the swapping interaction between the magnons and the photons, magnon mediated photon-phonon entanglement is generated. This entanglement can be nonreciprocally enhanced with taking both the Sagnac and the magnon Kerr effects into account. We also show that the achieved nonreciprocal entanglement can be kept up to $\sim3$ K with accessible parameters, exhibiting great potential for robustness against the bath temperature. Our paper provides a promising way to engineer various nonreciprocal devices with the magnon Kerr and the Sagnac effects in cavity magnomechanics.

\section{ ACKNOWLEDGMENTS}
This work is supported by the National Natural Science Foundation of China (Grant No. 12175001 ), the Natural Science Foundation of Zhejiang Province (Grant No. LY24A040004), and the key program of the Natural
Science Foundation of Anhui (Grant No. KJ2021A1301).

\bibliography{ms}
\end{document}